\documentclass[aps,pre,showpacs,twocolumn,floatfix]{revtex4}
\usepackage{graphics,epsfig,subfigure}
\usepackage{amsmath,amsfonts,amssymb,graphicx,longtable,dcolumn}

\begin{document}
\title{Thermodynamic stability of small-world oscillator networks:\\ A case study of proteins}

\author{Jie Ren}\email{renjie@nus.edu.sg}
\affiliation{NUS Graduate School for Integrative Sciences and
Engineering, Singapore 117597, Republic of Singapore}
\affiliation{Department of Physics and Centre for Computational
Science and Engineering, National University of Singapore, Singapore
117542}
\author{Baowen Li}\email{phylibw@nus.edu.sg}
\affiliation{NUS Graduate School for Integrative Sciences and
Engineering, Singapore 117597, Republic of Singapore}
\affiliation{Department of Physics and Centre for Computational
Science and Engineering, National University of Singapore, Singapore
117542}
\date{\today}

\begin{abstract}
We study vibrational thermodynamic stability of small-world
oscillator networks, by relating the average mean-square
displacement $S$ of oscillators to the eigenvalue spectrum of the
Laplacian matrix of networks. We show that the cross-links suppress
$S$ effectively and there exist two phases on the small-world
networks: 1) an unstable phase: when $p\ll1/N$, $S\sim N$; 2) a
stable phase: when $p\gg1/N$, $S\sim p^{-1}$, \emph{i.e.}, $S/N\sim
E_{cr}^{-1}$. Here, $p$ is the parameter of small-world, $N$ is the
number of oscillators, and $E_{cr}=pN$ is the number of cross-links.
The results are exemplified by various real protein structures that
follow the same scaling behavior $S/N\sim E_{cr}^{-1}$ of the stable
phase. We also show that it is the ``small-world" property that
plays the key role in the thermodynamic stability and is responsible
for the universal scaling $S/N\sim E_{cr}^{-1}$, regardless of the
model details.

\end{abstract}

\maketitle

Vibrational dynamics has been widely used to study thermodynamic
properties of various structures in solid state physics and/or
other disciplines \cite{Kittel}. Since the structure is considered
as a primary factor responsible for physical properties, to keep
the underlying structure thermally stable is of primary important
for systems to function properly. For example, proteins,
comprising an extremely heterogeneous class of biological
macromolecules, must be stable enough against thermal fluctuations
and/or external perturbations so as to maintain their native
structures and to function correctly \cite{Burioni2,Coex}.
Therefore, a natural and important question is often asked:  what
is the structure effect on thermodynamic stability? In this paper,
we study the stability of small-world structures \cite{WS}, which
is then exemplified by proteins.

The dynamics of $N$ coupled oscillators on the network in contact
with the external heat reservoir can be expressed as:
\begin{equation}
M\ddot{q}= -\sigma Lq-\Gamma\dot{q}+ \xi
\end{equation}
where $q=[q_1,q_2,...,q_N]^T$, denotes the oscillator's
displacements from the equilibrium positions.
$M_{ij}=m_i\delta_{ij}$ is the mass matrix, where $m_i$ denotes the
mass of the $i$th oscillator. $\sigma$ is the spring constant.
$L_{ij}=\delta_{ij}\sum_{m}A_{im}-A_{ij}$ is the Laplacian matrix
and $A_{ij}$ is the adjacency matrix of the network, where
$A_{ij}=1$ if $i$ and $j$ are connected and $A_{ij}=0$ otherwise.
$\Gamma_{ij}=\gamma_i\delta_{ij}$ is the dissipation matrix where
$\gamma_i$ is the dissipation coefficient of the $i$th oscillator
influenced by the heat reservoir. Vector
$\xi=[\xi_1,\xi_2,...,\xi_N]^T$ denotes the thermal fluctuation with
zero mean and variance $\langle \xi_i(t) \xi_j(t') \rangle =
2k_BT\Gamma_{ij}\delta(t-t')$, which is the usual
fluctuation-dissipation relation.

The harmonic potential we adopt looks very simple but can capture
the main features of the system. For example, Tirion \cite{Tirion}
demonstrates that a single-parameter harmonic potential can
reproduce vibrational properties of the real macromolecular system
very well. Thereafter, the Gaussian network model (GNM)
\cite{GNM1GNM2GNM3} has been widely used in protein research and
yields results in good agreement with experiments. In the GNM
model, the interactions are considered as homogeneous harmonic
springs, which is in analogy with the elasticity theory of random
polymer networks \cite{Flory, Pearson}.

\begin{figure*}
\scalebox{0.88}[0.88]{\includegraphics{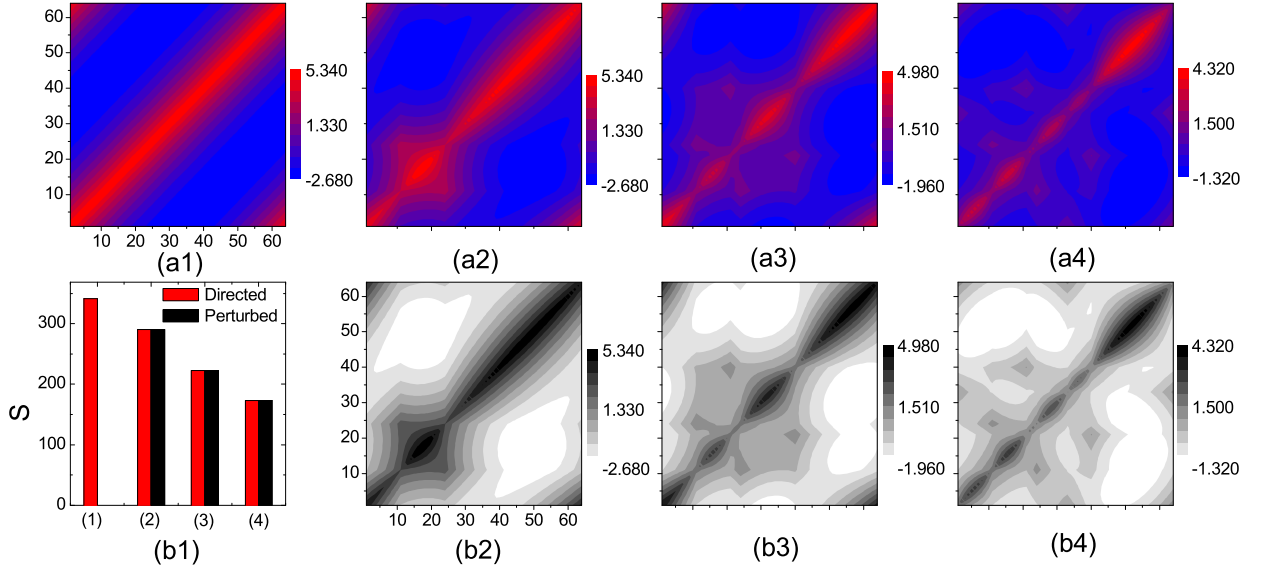}}
\caption{\label{fig:wide}(a1) The numerical result of the
correlation of pairwise oscillator displacements on a ring chain
$N=64$. (a2) The numerical result of the correlation for adding link
(11, 24) on (a1). (a3) is the numerical calculation of correlation
for adding link (21, 40) on (a2). (a4) is the correlation for adding
link (33, 64) on (a3). (b2), (b3), (b4) are the perturbed
calculation results using Eq. (8), corresponding to their
counterparts in the upper panel. (b1) shows the decreasing of $S$
with adding cross-links. The red ones are calculated directly from
numerical diagonalization and the black ones are the perturbed
calculation using Eq. (8). All the results validate our analytic
results. }
\end{figure*}

The correlation matrix of oscillator displacements at the steady
state for Eq. (1) can be easily obtained (see Appendix
\ref{app:A}):
\begin{equation}\label{2}
C_{lk}=\langle
q_lq_k\rangle=\frac{k_BT}{\pi}\int^{+\infty}_{-\infty}d\omega[G^{-1}(i\omega)\Gamma
G^{-1}(-i\omega)]_{lk},
\end{equation}
where matrix $G(\pm i\omega)=(\pm i\omega)^2M+(\pm
i\omega)\Gamma+\sigma L$. Since $G(i\omega)-G(-i\omega)=2i\omega
\Gamma$ and $G(0)=\sigma L$, one can eliminate $\Gamma$ in the
above integral and obtain:
\begin{equation}
C=-\frac{k_BT}{\pi
i}\int^{+\infty}_{-\infty}\frac{d\omega}{\omega}G^{-1}(i\omega)=\frac{k_BT}{\sigma}L^{\dagger},
\label{eq:C}\end{equation} where $L^{\dagger}$ denotes the
pseudo-inverse of $L$. It excludes zero mode which correspondes to
the translational invariance of the system, and is the inverse of
$L$ in the subspace orthogonal to the zero mode:
\begin{equation}
L^{\dagger}_{ij}=\sum_{\alpha=1}^{N-1} \frac{1}{\lambda_{\alpha}}
\psi_{\alpha i} \psi_{\alpha j},
\end{equation}
where $\lambda_{\alpha}$ are the non-zero eigenvalues, and
$\psi_{\alpha j}$ denote the corresponding normalized eigenvectors
of $L$. Therefore, we can obtain the average mean-square
displacement straightforwardly:
\begin{eqnarray} S = \frac{1}{N} \sum^{N}_{i}\langle
q_i^2\rangle = \frac{1}{N}\operatorname{tr}C
=\frac{k_BT}{N\sigma} \sum_{\alpha=1}^{N-1}
\frac{1}{\lambda_{\alpha}}. \label{eq:S}\end{eqnarray} This formula
relates the dynamic vibration property $S$ to the static structure
property --- the eigenvalue spectrum of Laplacian matrix $L$. When
the average mean-square displacement $S$ reaches the square of the
typical spacing between oscillators, the structure encounters large
vibrations and becomes unstable. Thus, small value of $S$ means
stable, while large value means unstable. It is clear that $S$ has a
trivial dependance on $T$ and $\sigma$ so that lower temperature or
larger spring constant indicates more thermal stability. Therefore,
to study the structure effect on $S$, we take $k_BT/\sigma=1$ in the
following, without loss of generality.

Based on Eq. (\ref{eq:C}) and (\ref{eq:S}), we can use exact
numerical diagonalizations of the Laplacian matrix $L$ to study
the structure effect on thermodynamic stability.

In fact, the stability can be also studied by perturbation analysis,
through which we find that the cross-links can suppress the
thermodynamic instability, \emph{i.e.,} decrease $S$ effectively.
After structure changes, the new Laplacian matrix is constructed as
$L'=L - \Delta$, where $\Delta$ denotes the perturbation. The new
correlation
matrix $C'$ can be written as 
\begin{equation}
C'=(I-C\Delta)^{-1}C=C + C\Delta C + C\Delta C \Delta C + \cdots.
\end{equation}
This is a standard algebraic algebra treatment of matrix
perturbation, which can be simply regarded as Taylor series. In
fact, it is similar to the Dyson equation in Feynman-Dyson
perturbation theory \cite{dyson}. For the case of adding a link
between nodes $i$ and $j$, the perturbation matrix $\Delta$ can be
expressed as
\begin{equation}
\Delta_{mn}=\delta_{mi}\delta_{nj}+\delta_{mj}\delta_{ni}-\delta_{mi}\delta_{ni}-\delta_{mj}\delta_{nj}.
\end{equation}
Substituting the expression of $\Delta$ into $C'$, one obtains
\begin{equation}
C'_{mn}=C_{mn}-\frac{(C_{mi}-C_{mj})(C_{in}-C_{jn})}{1+R_{ij}},
\end{equation}
where
$R_{ij}=C_{ii}+C_{jj}-2C_{ij}=\sum^{N-1}_{\alpha=1}(\psi_{\alpha
i}-\psi_{\alpha j})^2/\lambda_{\alpha}$. Therefore, the new average
mean-square displacement is
\begin{eqnarray}
S' &=& \frac{1}{N}\operatorname{tr}C' 
= S-\frac{\sum^{N}_{k=1}(C_{ik}- C_{jk})^2}{N(1+R_{ij})} \nonumber \\
&=& S-\frac{1}{N(1+R_{ij})}\sum^{N}_{k=1}\frac{(\psi_{\alpha
i}-\psi_{\alpha j})^2}{\lambda^2_{\alpha}}.
\label{eq:perturbation}\end{eqnarray} The second term in Eq.
(\ref{eq:perturbation}) is positive so that the value of new $S'$ is
always smaller than $S$. In other words, the cross-links always
decrease $S$ so as to increase the thermodynamic stability of the
system. A specific case is studied and illustrated in Fig.
\ref{fig:wide}.

\begin{figure}
\scalebox{0.4}[0.4]{\includegraphics{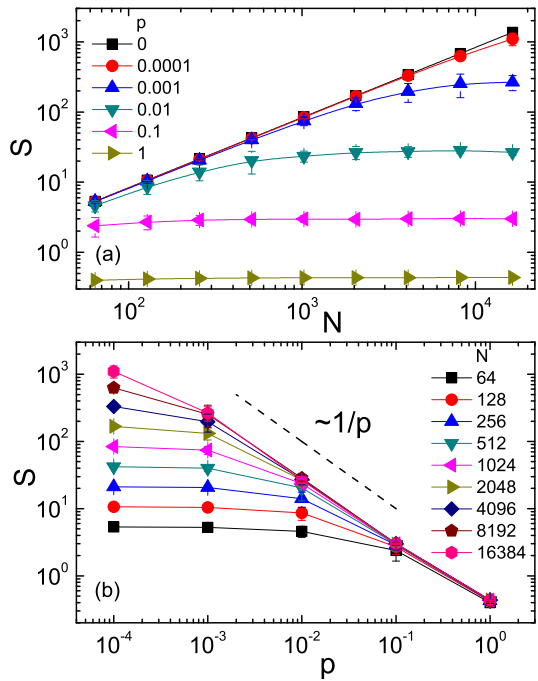}}
\caption{\label{fig:s-n}(color online). (a) $S$ versus network size
$N$. The straight line for $p=0$ indicates the diverging behavior,
$S\sim N$, of the 1D ring chain. Even small nonzero value of $p$ can
suppress the diverging behavior of $S$ and make it saturated to a
finite value in the large size limit. (b) For each $N$, $S$
decreases as $p$ increases. A scaling behavior $S\sim 1/p$ emerges
in the large size limit. All the results indicate that the
cross-links boost the thermodynamic stability effectively.}
\end{figure}

In the following study, we choose a typical model to construct the
small-world structure \cite{sw}. We first consider $N$ oscillators
(which might be an atom, a molecule, or other module structure,
depending on the system studied) on a one-dimensional(1D) ring
chain, \emph{i.e.}, with periodic boundary conditions. Each
oscillator is connected to its nearest-neighbors. Then, we add a
cross-link to each oscillator with probability $p$, which connects
to another non-neighboring oscillator randomly. Thus, $E_{cr}=pN$ is
the number of cross-links. When $p=0$, the structure reduces to the
1D ring chain. In all cases studied below, each data point is
obtained by averaging over $50$ different network configurations for
a given $p$ and $N$.

Figure \ref{fig:s-n}(a) illustrates $S$ versus the system size $N$
for different values of $p$ in double logarithmic scale. The $p=0$
case, corresponding to the 1D ring structure, shows the power-law
divergence, $S\sim N$. It indicates that no thermodynamically stable
solid exists at finite temperature in 1D. Indeed, when the average
mean-square displacement $S$ exceeds the square of the typical
spacing between oscillators, the structure behaves like a liquid
rather than a solid, and the crystalline order makes no sense
anymore. This behavior is also reported in \cite{Pierels, Burioni1}.
For the case of $p\neq0$, even of small value, as $N$ increases, $S$
is saturated to a finite value rapidly. Moreover, from Fig.
\ref{fig:s-n}(b), we can see that the larger $p$, the smaller $S$
and in the large $N$ limit, a scaling $S\sim p^{-1}$ emerges. All
the above results indicate that the cross-links suppress the average
mean-square displacement $S$ effectively and make it convergent in
the thermodynamic limit.

\begin{figure}
\scalebox{0.48}[0.48]{\includegraphics{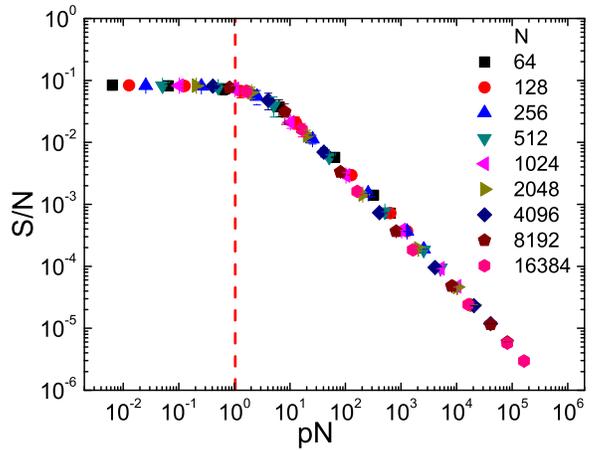}}
\caption{\label{fig:scaling}(color online). Scaling plot of the
average mean-square displacement $S$ in small-world networks for
various size $N$ and probability $p$. All data are from Fig.
\ref{fig:s-n} \cite{bigP} and they collapse into one single line
very well. It shows two phases clearly: one is the regime with
slope $-1$, where $p\gg1/N$,  indicating the non-divergent stable
behavior $S/N\sim E_{cr}^{-1}=(pN)^{-1}$, \emph{i.e.}, $S\sim
1/p$. Another one is the horizontal regime, indicating the
diverging unstable behavior, $S\sim N$. The red vertical dashed
line is used for eye-guiding to separate the two phases. }
\end{figure}

To eliminate the finite size effect, all the data from Fig.
\ref{fig:s-n} \cite{bigP} are re-scaled and the results are
illustrated in Fig. \ref{fig:scaling}. It is found that all data
points collapse into one single line very well and two distinct
phases emerge. For $p\ll 1/N$, there is a horizontal regime,
$S/N\sim$ const. It indicates an unstable phase: when the number of
cross-links is smaller than one, $S$ diverges with $N$. For
$p\gg1/N$, there is a regime with slope $-1$, $S/N\sim
E_{cr}^{-1}=(pN)^{-1}$, which indicates a convergent stable phase.
In other words, when the number of cross-links is much larger than
one, $S$ approaches a finite value at large $N$ and scales as
$p^{-1}$.

Since the average mean-square displacement $S$ is related to the
spectral properties of the Laplacian matrix $L$, we can understand
the scaling behavior of $S$ in terms of its eigenvalue spectrum
$\rho(\lambda)$. For large size $N$, Eq. (\ref{eq:S}) can be
expressed as  \cite{Burioni1}:
\begin{equation}
S=\int\frac{\rho(\lambda)}{\lambda}d\lambda,
\label{eq:S2}\end{equation} from which we can easily see that the
density of small $\lambda$ dominates the behavior of $S$. For the
case without cross-links, the system reduces to a 1D ring chain,
where $\rho(\lambda)\sim \lambda^{-1/2}$ and $\lambda\sim N^{-2}$
for small $\lambda$. Thus, $S\sim\int\lambda^{-3/2}d\lambda\sim N$.
For the case with cross-links, following the heuristic argument in
\cite{BrayMonasson}, we can consider that the ring chain is divided
into several quasi-linear segments of length $l$, and the
probability of length $l$ is exponentially small, $e^{-pl}$. Each
segment $l$ contributes to small eigenvalues of the order of
$l^{-2}$. Summing over lengths with the exponential weight, we
obtain
$S=\int\frac{\rho(\lambda)}{\lambda}d\lambda\sim\int^{N}_{0}\frac{l^{-2}e^{-pl}}{1/l^2}dl
=\frac{1}{p}(1-e^{-pN})$. When $pN\ll1$, $S\sim N$; while $pN\gg1$,
$S\sim 1/p$, which is exactly what our numerical results show in
Fig. \ref{fig:s-n} and \ref{fig:scaling}. Although the argument
above is not rigorous and applies only when $p$ is smaller than one,
it gives us quite good understanding of the scaling behavior of $S$.
When $p$ is larger than one, the model we used is more like an
Erd\"{o}s-R\'{e}nyi model, which is also to be demonstrated to
follow the same scaling of the stable regime at the end of this
paper.

The small-world structure we used above is well studied \cite{sw}.
Using renormalization group method, the authors in Ref. \cite{sw}
showed that this model undergoes a transition between regular
lattice and random one at intermediate characteristic size
$N_c\sim p^{-1}$. In other words, the phase transition has a
critical point $p_c=0$ in the thermodynamical limit when
$N\rightarrow\infty$. For finite size $N$, the diameter $l$ scales
linearly with $N$ for $N_c \gg N$ as it is in 1D ring chain, while
$l\sim\ln{N}$ for $N \gg N_c$, where it exhibits ``small-world"
property. Our results about unstable and stable phases are
consistent with their findings that the unstable regime
corresponds to the 1D case and the stable phase corresponds to the
``small-world" case.

As an illustrative example, the thermodynamic stability is further
tested on real protein data. We revisit the proteins used in Ref.
\cite{Burioni2}, which differ in functions and structures, with a
wide size scale ranging from $100$ to $3600$ residues. All the
structure data are downloaded from the Protein Data Bank (PDB)
\cite{PDB}
 and the number of all residue pairs is counted within a customary
cutoff $7.0$ \AA. After eliminating the connectivity number of the
primary structure of protein from the counted number, we obtain
the number of cross-links, $E_{cr}$, for each protein. The mean
square displacement of C$^{\alpha}$ atoms is characterized by
$B$-factor \cite{Bfactor}, also called Debye-Waller or temperature
factor: $B_i=8\pi^2\langle q^2_i\rangle/3$, where $i$ is the index
of amino acid residue. It is experimentally measured via x-ray
crystallography, and also can be download from the PDB.  The
average $B$-factor is calculated over all C$^{\alpha}$ atoms for
each protein, $B=\sum_{i=1}^{N}B_i/N$. Notice that at above
theoretical analysis, $k_BT/\sigma$ is set to be $1$ for
convenience, which is not always true. The value of $k_BT/\sigma$
varies among different proteins. Thus, we use the estimated data
of $k_BT/\sigma$ \cite{Burioni2} to obtain the normalized average
$B$-factor, $B^{\prime}=B/(k_BT/\sigma)$. Note that $B^{\prime}$
is analogous to $S$, defined in Eq. (\ref{eq:S2}). All the details
of these proteins are listed in Table \ref{table} in Appendix
\ref{app:B}.

\begin{figure}
\scalebox{0.43}[0.43]{\includegraphics{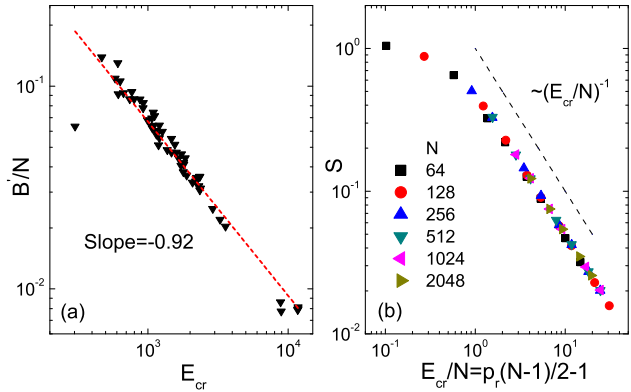}}
\caption{\label{fig:pdb}(color online). (a) Log-log plot of
normalized average $B$-factor with various number of cross-links
for real protein data. It exhibits clearly a power-law behavior,
$B^{\prime}/N\sim E_{cr}^{-a}$. The dashed line indicates the
best-fit of the power-law, with exponent $a=-0.92\pm0.01$. (b) The
average mean-square displacement $S$ versus the parameter
$E_{cr}/N=p_r(N-1)/2-1$ in Erd\"{o}s-R\'{e}nyi model. The dashed
line indicates the scaling.}
\end{figure}

The thermodynamic stability is crucial to keep the native structure
of protein for right function. Moreover the structure of protein is
also found to have ``small-world" property \cite{swprotein12},
\emph{i.e.}, $l\sim\ln{N}$. It is intuitive for us to expect that
nature selection forces proteins evolving into the stable phase in
Fig. \ref{fig:scaling}, which implies $B^{\prime}/N\sim
E_{cr}^{-1}$. Figure \ref{fig:pdb}(a) verifies our expectation drawn
from the argument of stability analysis. In fact, we obtain a clear
power-law
scaling: 
\begin{equation}
B^{\prime}/N\sim E_{cr}^{-a}, \ a=0.92\pm0.01,
\end{equation}
which is quite close to $1$. This scaling reveals the universal
behavior shared by various different proteins, regardless of their
sources or functions. It implies an underlying general mechanism
that nature selects proteins with thermodynamic stability
constraints.

\begin{figure}
\scalebox{0.44}[0.44]{\includegraphics{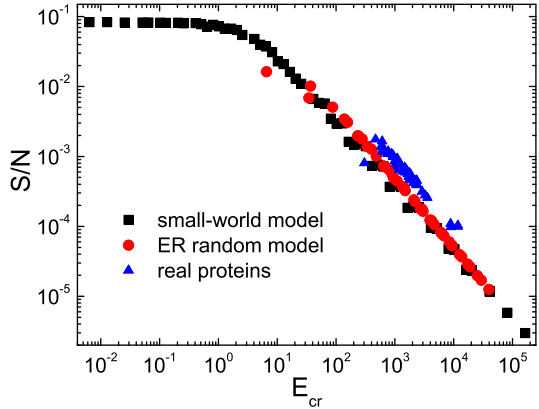}} \caption{(color
online). $S/N$ versus $E_{cr}$ for three different networks. It
shows that the ``small-world" property is responsible for the
universal scaling $S/N\sim E_{cr}^{-1}$ in the stable regime,
regardless of the model details. Note that for proteins,
$S=B'/(8\pi^2)$, where the factor $3$ is removed since $B$-factor is
measured in $3$-dimension.}
\end{figure}

Although protein has more complex structures with high
modularity(domains), ``small-world" property captures its main
feature. Thus, ``small-world" might play a key role in the
thermodynamic stability of structures and be responsible for the
scaling in the stable regime. To validate our conjecture, we
further study the thermodynamic stability in
Erd\"{o}s-R\'{e}nyi(ER) random network model in the following.

ER model \cite{bollobas} has $N$ nodes and every pair of nodes is
connected with probability $p_r$. The average degree $\langle
k\rangle=p_r(N-1)$. There are several phases in this model depending
on different threshold $p_r$: when $\langle k\rangle=p_r(N-1)>3.5$
\cite{bollobas}, the diameter of the graph equals the diameter of
the giant cluster, and is proportional to $\ln{N}$, \emph{i.e.},
``small-world" property. Thus, it is straightforward to expect that
ER model might share the same behavior $S/N\sim E_{cr}^{-1}$. The
numerical result is illustrated in Fig. 4(b). As we point out above,
ER model has ``small-world" property \cite{explain} when
$p_r(N-1)>3.5$. Correspondingly, when $p_r(N-1)/2-1>0.75$, $S\sim
N/E_{cr}=N/(p_rN(N-1)/2-N)=[p_r(N-1)/2-1]^{-1}$ as shown in Fig.
4(b).

For convenience of comparison, we plot the data of three cases
together in Fig. 5. It clearly shows the universal scaling $S/N\sim
E_{cr}^{-1}$ in the regime where the three structures all have
``small-world" property, $l\sim\ln N$. Moreover, we have tested
other network models \cite{scalefree} sharing the property $l\sim\ln
N$. The results indicate that the ``small-world" property plays the
key role in the stable regime and is responsible for the universal
scaling, regardless of the model details, which can be explained in
the framework of a mean-field approach \cite{scalefree}.

In summary, we have studied the vibrational thermodynamic stability
of small-world structures. The average mean-square displacement $S$
of the structure has been expressed as the mean of inverse
eigenvalues of its Laplacian matrix $L$. Therefore, the dynamic
vibration property is closely related to the static structure
information. It is found that the cross-links suppress $S$
effectively and on the small-world network model, there exist two
phases: an unstable phase where $p\ll1/N$, $S\sim N$ and a stable
phase where $p\gg1/N$, $S\sim p^{-1}$, \emph{i.e.}, $S/N\sim
E_{cr}^{-1}$. Further, we have tested various data from the PDB and
find that native proteins belong to the stable phase and share the
same scaling behavior $S/N\sim E_{cr}^{-1}$. It is believed that
nature selects proteins under the constraint of thermodynamic
stability so that proteins can keep their specific native fold
structure stable for proper function. Finally, we have studied $S$
in ER random network model and have validated our conjecture that it
is the ``small-world" property that plays a key role in the
thermodynamic stability of structures and is responsible for the
universal scaling, $S/N\sim E_{cr}^{-1}$, in the stable regime. It
is also interesting to examine more complex structure effects on the
thermodynamic stability problem, such as scale-free networks
\cite{scalefree}, hierarchical structures, networks with community
structure \emph{etc}. More realistic considerations such as the
effect of random coupling constants, anharmonic potentials, or even
quantum version of vibration dynamics are worth further studying.

The work is supported by the NUS Faculty Research Grant No.
R-144-000-165-112/101.

\appendix

\section{Derivation of the correlation matrix} \label{app:A}

To make this paper self-contained and readable, we complement the
detailed derivation of Eq. (\ref{2}) which is expressed in terms
of $G$ in Fourier transform space. We follow Ref. \cite{kubo} by
defining the Fourier transform as

\begin{eqnarray}
Q(\omega)&=&\frac{1}{2\pi}\int^{+\infty}_{-\infty}q(t)e^{-i\omega
t}dt;
\\
\eta(\omega)&=&\frac{1}{2\pi}\int^{+\infty}_{-\infty}\xi(t)e^{-i\omega
t}dt. 
\end{eqnarray}

Applying Fourier transform to both sides of Eq. (1), one obtains
\begin{equation}
-\omega^2MQ= -\sigma LQ-i\omega\Gamma Q+ \eta. 
\end{equation}

Simple algebraic operation yields,
\begin{equation}
Q=G^{-1}(i\omega)\eta, 
\end{equation}
where matrix $G(i\omega)=-\omega^2M+i\omega\Gamma+\sigma L$ as
defined in text. The two point correlation function is
\begin{eqnarray}\label{16}
&&\langle q_l(t+\tau)q_k(t)\rangle\nonumber
\\&=&\int^{+\infty}_{-\infty}d\omega
e^{i\omega(t+\tau)}\int^{+\infty}_{-\infty}d\omega'e^{-i\omega'
t}\langle Q_l(\omega)Q_k^{*}(\omega')\rangle \nonumber \\
&=&\int^{+\infty}_{-\infty}d\omega
e^{i\omega(t+\tau)}\int^{+\infty}_{-\infty}d\omega'e^{-i\omega'
t}\nonumber
\\&&\times \langle \eta(\omega)\eta^{*}(\omega')\rangle
[G^{-1}(i\omega)G^{-1}(-i\omega')]_{lk}. 
\end{eqnarray}
where $*$ denotes the conjugate transpose and
$Q^{*}(\omega')=\eta^{*}(\omega')G^{-1}(-i\omega')$. Moreover,
since
\begin{eqnarray}\label{17}
\langle
\eta(\omega)\eta^{*}(\omega')\rangle&=&\frac{1}{2\pi}\int^{+\infty}_{-\infty}
dte^{-i\omega t}\nonumber \\&&\times
\frac{1}{2\pi}\int^{+\infty}_{-\infty}dt'e^{i\omega'
t'}\langle\xi(t)\xi^{*}(t')\rangle \nonumber \\
&=&\frac{k_BT\Gamma}{\pi}\frac{1}{2\pi}\int^{+\infty}_{-\infty}
e^{i(\omega'-\omega)t} \nonumber \\
&=&\frac{k_BT\Gamma}{\pi}\delta(\omega'-\omega), 
\end{eqnarray}
substitute Eq. (\ref{17}) into Eq. (\ref{16}) and we have:
\begin{eqnarray}
&&\langle q_l(t+\tau)q_k(t)\rangle \nonumber
\\&=&\frac{k_BT}{\pi}\int^{+\infty}_{-\infty}d\omega
e^{i\omega\tau}[G^{-1}(i\omega)\Gamma G^{-1}(-i\omega)]_{lk}.
\end{eqnarray}
The expression of correlation matrix Eq. (\ref{2}) corresponds to
the special case $\tau=0$.

\section{Information of Proteins} \label{app:B}
\begin{longtable*}
 {m{45pt} m{25pt} m{30pt} m{40pt} m{35pt} m{35pt}|m{45pt} m{25pt} m{30pt} m{40pt} m{35pt} m{35pt}}\hline\hline
 PDB code   & $N$  & $E_{cr}$   & $B$ & $k_BT/\sigma$  & $B'/N$ & PDB code   & N  & $E_{cr}$  & $B$ & $k_BT/\sigma$  & $B'/N$\\ \hline
 9RNT   & 104   & 303   & 10.9147  &    1.657  &   0.06334 & 16PK   & 415   & 1472   & 14.3769   & 0.63   & 0.05499 \\
 1BVC   & 153   & 469   & 8.33124   & 0.392   &   0.13891 & 1BU8   & 446   & 1559   & 19.7459   & 0.859   & 0.05154 \\
 1G12   & 167   & 584   & 14.4393   & 0.793   & 0.10903 & 1AC5   & 483   & 1598   & 24.8104  & 1.091   & 0.04708 \\
 1AMM   & 174   & 612   & 0.06793   & 0.003   & 0.13013 & 1LAM   & 484   & 1737   & 10.9112   & 0.488   & 0.0462 \\
 1KNB   & 186   & 616   & 18.7711   & 1.104   & 0.09141 & 1CPU   & 495   & 1659   & 13.5504   & 0.62   & 0.04415 \\
 1CUS   & 197   & 671   & 16.6598   & 0.914   & 0.09252 & 3COX   & 500   & 1792   & 9.2601   & 0.491   & 0.03772 \\
 1IQQ   & 200   & 634   & 10.164   & 0.48   & 0.10588 & 1A65   & 504   & 1724   & 21.3040   & 1.042   & 0.04057 \\
 2AYH   & 214   & 744   & 9.9678   &  0.539   & 0.08642 & 1SOM   & 528   & 1805   & 34.3889   & 1.585   & 0.04109 \\
 1AE5   & 223   & 768   & 19.9342   & 0.952   & 0.09390 & 1E3Q   & 534   & 1799   & 35.0785   & 1.577   & 0.04181 \\
 1LST   & 239   & 799   & 20.2462   & 0.982   & 0.08627 & 1CRL   & 534   & 1893   & 18.7736   & 0.969   & 0.03628 \\
 1A06   & 279   & 880   & 52.5323   & 2.184   & 0.08621 & 1AKN   & 547   & 1851   & 41.9999   & 1.737   & 0.0442 \\
 1NAR   & 289   & 925   & 13.5809   & 0.602   & 0.07806 & 1CF3   & 581   & 2082   & 22.4561   & 1.154   & 0.03349 \\
 1A48   & 298   & 928   & 16.3599   & 0.664   & 0.08268 & 1EX1   & 602   & 2199   & 24.2479   & 1.193   & 0.03376 \\
 1A3H   & 300   & 1076   &13.4101   & 0.719   & 0.06217 & 1A14   & 612   & 2198   & 17.9869   & 0.865   & 0.03398 \\
 1SBP   & 309   & 1061   & 12.5878   & 0.641   & 0.06355 & 1MZ5   & 622   & 2234   & 16.4778   & 0.75   & 0.03532 \\
 1A5Z   & 312   & 1070   & 45.6872   & 2.111   & 0.06937 & 1CB8   & 674   & 2348   & 28.0511   & 1.164   & 0.03575 \\
 1A1S   & 313   & 1088   & 21.3477   & 1.068   & 0.06386 & 1HMU   & 674   & 2341   & 21.8915   & 0.907   & 0.03581 \\
 1ADS   & 315   & 993    & 10.7205   & 0.5   & 0.06807 & 1A47   & 683   & 2350   & 13.5361   & 0.646   & 0.03068 \\
 1A40   & 321   & 1153   & 9.72025   & 0.524   & 0.05779 & 1CDG   & 686   & 2375   & 23.1041   & 1.074   & 0.03136 \\
 1A54   & 321   & 1144   & 11.6098   & 0.601   & 0.06018 & 1DMT   & 696   & 2313   & 26.9702   & 1.204   & 0.03218 \\
 1A0I   & 332   & 1094   & 27.2887   & 1.109   & 0.07412 & 1A4G   & 780   & 2904   & 11.4584   & 0.591   & 0.02486 \\
 3PTE   & 347   & 1210   & 8.13032   & 0.366   & 0.06402 & 1HTY   & 988   & 3276   & 14.0281   & 0.646   & 0.02198 \\
 1A26   & 351   & 1117   & 34.2736   & 1.369   & 0.07133 & 1KCW   & 1017   & 3579   & 44.0269   & 2.13   & 0.02032 \\
 1BVW   & 360   & 1209   & 13.0188   & 0.652   & 0.05547 & 1KEK   & 2462   & 8860   & 26.7142   & 1.263   & 0.00859 \\
 8JDW   & 360   & 1191   & 23.8334   & 1.293   & 0.05120 & 1B0P   & 2462   & 8936   & 6.08348   & 0.319   & 0.00775 \\
 7ODC   & 387   & 1266   & 19.8278   & 0.859   & 0.05964 & 1K83   & 3490   & 11725   & 55.9118   & 2.03   & 0.00788 \\
 1OYC   & 399   & 1378   & 20.4127   & 1.056   & 0.04845 & 1I3Q   & 3542   & 11813   & 70.0899   & 2.435   & 0.00813 \\
 1A39   & 410   & 1474   & 21.4742   & 1.113   & 0.04706 & 1I50   & 3558   & 11799   & 63.2545   & 2.236   & 0.00795 \\ \hline\hline
\caption{\label{table} Information of Proteins used in the present
study. Size $N$ is the number of residues. $E_{cr}$ is the number
of cross-links, counted within cutoff 7 \AA. $B$ is the average
$B$-factor over all C$^{\alpha}$ atoms for each protein. The
estimated $k_BT/\sigma$ are collected in Ref. \cite{Burioni2}.
$B^{\prime}/N$ is the normalized $B$-factor over the size of
protein.}
\end{longtable*}


\begin{thebibliography}{10}
\bibitem{bigP} We add two more cases $(p=5$ and $p=10)$ for all sizes. Here, $p$ is larger than one, which means adding to each oscillator $p$ cross-links on average.
\end{thebibliography}
\end{document}